\documentclass[aps,prl,twocolumn,showpacs]{revtex4-1}
\usepackage{graphicx,amsmath,amssymb,bm}

\usepackage[usenames]{color}

\begin{document} 
 
\title{Non-perturbative Extraction of the Effective Mass in Neutron Matter}

\author{Mateusz Buraczynski}
\affiliation{Department of Physics, University of Guelph, Guelph, ON N1G 2W1, Canada}
\author{Nawar Ismail}
\affiliation{Department of Physics, University of Guelph, Guelph, ON N1G 2W1, Canada}
\author{Alexandros Gezerlis}
\affiliation{Department of Physics, University of Guelph, Guelph, ON N1G 2W1, Canada}

\begin{abstract} 
We carry out non-perturbative calculations of the single-particle
excitation spectrum in strongly interacting neutron matter. These are 
microscopic quantum Monte Carlo computations of many-neutron energies at
different densities as well as several distinct excited states.
As input, we employ both phenomenological and chiral two- and three-nucleon interactions.
We use the single-particle spectrum to extract the effective mass in neutron matter.
With a view to systematizing the error involved in this extraction, we 
carefully assess the impact of finite-size effects on the quasiparticle dispersion relation. We find an effective-mass
ratio that drops from 1 as the density is increased.
We conclude by connecting our results with the physics of ultracold gases
as well as with 
energy-density functional theories of nuclei and neutron-star matter. 
\end{abstract} 

 
\maketitle

The physics of neutron matter (NM) is directly related to the properties of neutron stars~\cite{Gandolfi:2015}.
Following several decades of \textit{ab initio} work~\cite{Friedman:1981,Akmal:1998,Schwenk:2005,Gezerlis:2008,Epelbaum:2008b,Kaiser:2012}, pure neutron systems have also served as the natural testing grounds of nuclear forces, whether 
phenomenological~\cite{Carlson:Morales:2003,Gandolfi:2009,Gezerlis:2010,Gandolfi:2012,Baldo:2012} 
or chiral~\cite{Hebeler:2010,Gezerlis:2013,Coraggio:2013,Hagen:2014,Gezerlis:2014,Carbone:2014,Roggero:2014,Wlazlowski:2014,Soma:2014,Tews:2016,Piarulli:2018,Lonardoni:2018}.
Today, after the recent detection of a gravitational-wave signal from a neutron-star merger~\cite{LIGO,Tews:2018}, 
the field of neutron-rich matter
has entered a new era, where microscopic predictions will foreseeably be confronted with experimental measurements.
The connection between pure-neutron calculations and the properties of neutron-rich nuclei involves 
the use of nuclear energy-density
functionals (EDFs): these currently constitute the only approach that is able to globally describe the  
nuclear chart~\cite{Bender:2003}. EDFs and \textit{ab initio} many-body calculations have a long
history of fruitful interaction: typically, undetermined parameters in EDFs are fit either 
to nuclear masses and radii or to ``synthetic data'' coming from many-body calculations. 
Such many-body results employed as constraints follow from 
both homogeneous and inhomogeneous systems, ranging from the EOS of neutron matter~\cite{Fayans,SLy,Brown:2000,
Gogny,Fattoyev,Brown:2014,Rrapaj:2016},
to the neutron pairing gap~\cite{Chamel:2008}, 
the neutron polaron~\cite{Forbes:2014,Roggero:2015}, 
the setting of neutron drops~\cite{Pudliner:1996,
Pederiva:2004,Gandolfi:2011,Potter:2014}, or the static response problem~\cite{Pastore:2015,Chamel:2014,Davesne:2015,Buraczynski:2016,Buraczynski:2017,Boulet:2018}. 

In this Letter, we continue on this path of constraining 
phenomenology using \textit{ab initio} nuclear theory.
The goal here is to carry out first-principles calculations of the quasiparticle
energy dispersion relation in neutron matter and then try to use these to extract generally meaningful quantities.
Specifically, we focus on one of the most basic parameters used in EDFs, namely the effective mass
near the Fermi surface (see Ref.~\cite{Li:2018} for a comprehensive review). 
Effective masses are important because they can impact thermodynamic properties,
the maximum mass of a neutron star, the static response of nucleon matter,
as well as analyses of giant quadrupole resonances.
While analogous extractions of the effective mass have been carried out using other many-body
methods~\cite{Friedman:1981,
Wambach:1993,Schwenk:2003,Drischler:2014,Isaule:2016,Grasso:2018,Bonnard:2018}, this is the first time a 
controlled non-perturbative nuclear technique has been employed for this problem. 
Since related calculations have been compared to cold-atom experiment for balanced systems or for 
impurities~\cite{Carlson:2005,Lobo:2006,Ku:2012,Forbes:2014,Roggero:2015}, we will touch upon 
possible connections.

We start from a microscopic Hamiltonian
made up of a non-relativistic kinetic energy, a two-nucleon (NN) interaction  
and a three-nucleon (NNN) interaction. In other words:
\begin{equation}
\hat{H} = -\frac{\hbar^2}{2m}\sum_i \nabla^{2}_{i} +\sum_{i<j}V_{ij}+\sum_{i<j<k}V_{ijk}
\label{eq:ham}
\end{equation}
The NN and NNN interactions here
are taken from two families: a) high-quality phenomenology
(specifically, the Argonne $v8'$
potential~\cite{Wiringa:2002} and the Urbana IX potential~\cite{Pudliner:1997}),
and b) local chiral forces (specifically, the $R_0 = 1.0$ fm NN interaction of Ref.~\cite{Gezerlis:2014}
and the $R_{3N} = 1.0$ fm NNN interaction of Ref.~\cite{Tews:2016}). 
Qualitatively, we do not expect
the details of the interaction to have much of an impact at low density. At the level of
the equation-of-state~\cite{Gandolfi:2015,Tews:2016}, at larger densities the interaction
does start to play a quantitative role: it is not clear \textit{a priori} what the corresponding
effect will be at the level of the single-particle spectrum. This motivates our choice to 
compute the same quantities using both phenomenological and chiral potentials:
carrying out twice as many QMC calculations can help us understand 
the difference between bulk vs single-particle properties in neutron matter. 

To carry out our many-neutron computations, we use two successful \textit{ab initio} non-perturbative many-body 
approaches, which belong to the quantum Monte Carlo family~\cite{Pudliner:1997,Gandolfi:2012,Gezerlis:2013}.
Specifically, our trial wave function (including spins) 
has the form:
\begin{equation}
|\Psi_T \rangle =\prod_{i<j}f(r_{ij})\,\,\mathcal{A}\bigg[\prod_i|\phi_i,s_i\rangle\bigg]
\label{eq:trial}
\end{equation}
The first term is a nodeless Jastrow factor so it should, in principle, only impact the variance of the final 
answer. The second term is a Slater determinant of single-particle orbitals. Since we are interested in 
homogeneous infinite neutron matter, we take these to be plane waves: for a single-particle position $\mathbf{r}_i$,
the plane wave is $e^{i \mathbf{k}\cdot \mathbf{r}_i}$ and the $\mathbf{k}$ has to obey the Born-von Karman periodic
boundary conditions:
\begin{equation}
\mathbf{k} = \frac{2 \pi}{L} \left ( n_x, n_y, n_z \right )
\label{eq:Born}
\end{equation}
where $L$ is the simulation-box length and $n_x$, $n_y$, $n_z$ are integers. Since we're dealing with spin-1/2 fermions (neutrons),
we can have up to two different particles for a given choice of $n_x$, $n_y$, and $n_z$
(one for spin-up projection and the other for spin-down). At zero temperature for the many-neutron case,
the particles occupy the lowest available $k$ states. Since the many-neutron wave function is ambiguous for the case where 
a shell is not completely filled, we generally prefer to study closed shell configurations, found 
at $N = 2, 14, 38, 54, 66, 114, \ldots$. 

Let's take some time to examine the highest occupied momentum state:
we denote this by $k_{F, N}$ where $N$ is a given choice. When working at fixed
number density, $n = N/V = k_F^3/(3\pi^2)$, this $k_F$ is a thermodynamic-limit quantity (i.e., it corresponds to the limits
$N \rightarrow \infty$ and $V \rightarrow \infty$, while $n \rightarrow $ const).
When $N$ is finite but large, we expect $k_{F, N} \approx k_F$; as $N$ becomes small, it is reasonable to expect
deviations. 

As a first many-body calculational step, we use variational Monte Carlo (VMC) to produce starting configurations.
For the case of frozen spins, one then proceeds to employ diffusion Monte Carlo (DMC), which is 
an accurate method for computing the ground-state energy 
of a many-body system. Starting from a trial wave function $|\Psi_T \rangle$ as input, one projects 
out the excited states by evolving forward in imaginary time. 
Auxiliary Field Diffusion Monte Carlo (AFDMC) \cite{Schmidt:1999} extends DMC to the case of Hamiltonians with a complicated spin dependence. 
In essence, this method reduces the number of operations for handling 
 spin from exponential to linear at the cost of introducing a set of auxiliary fields. Since the floating-point operations 
 required for an AFDMC calculation scale as $N^3$, simulations are typically limited to roughly 100 particles or so.
 The DMC and AFDMC methods have been very successful in describing neutron matter from low- to high-density~\cite{Gandolfi:2015}.
  
In our calculations for many-neutron systems in this Letter, 
we study densities from $0.02$ to $0.20$ fm$^{-3}$. The lowest density
is large enough that pairing effects are not significant \cite{Gezerlis:2008}, whereas the highest density
is near the point where we no longer expect chiral forces to be dependable (chiral effective field
theory interactions result from a low-momentum expansion, after all). 
In QMC calculations of the equation-of-state of neutron matter~\cite{Gezerlis:2008,Gezerlis:2010,Gezerlis:2013}
it is standard to carry out simulations for 66 particles.

\begin{figure}[b]
\begin{center}
\includegraphics[width=1.0\columnwidth,clip=]{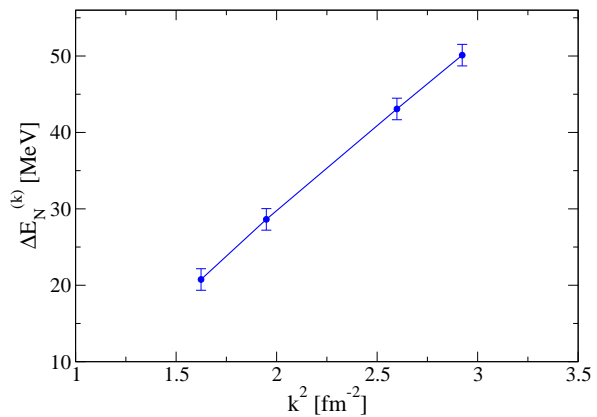}
\caption{(color online) Neutron-matter quasiparticle energy dispersion, $\Delta E^{(k)}_{N}$,
as a function of the square of the momentum, $k^2$, of the $67$-th particle (using AFDMC and AV8'+UIX
at a density of $0.05$ fm$^{-3}$). Since this is 
a (non-superfluid) Fermi system, we are not free to add the extra particle
at any momentum: thus, this is the energy dispersion near the Fermi surface.
The lines connecting the dots are there to guide the eye.
\label{fig:vsk}}
\end{center}
\end{figure}

As a first step in this study, we look at comparing the energies of the $N$-particle system and the $(N+1)$-particle system.
When taking the difference of these two energies, one has to be a little careful:
in a QMC calculation, one has to think about which parameter is kept fixed. If you carry out the $N$- and $(N+1)$-particle 
calculations keeping the volume constant, you pay the price of having the two energies be evaluated at slightly
different densities. If, instead, you choose to simulate the $N$- and $(N+1)$-particle systems at constant 
density, then you have to account for the energy loss due to the expanding box size. One can show that,
after introducing this density correction, the quasiparticle energies of the free and interacting systems 
can be expressed as follows:
\begin{align}
\Delta T^{(k)}_{N} \equiv T^{(k)}_{N+1} - T_N + \frac{2}{5} E_F &= \frac{\hbar^2 k^2}{2m} \nonumber \\
\Delta E^{(k)}_{N} \equiv E^{(k)}_{N+1} - E_N + \frac{2}{5} \xi E_F &= \frac{\hbar^2 k^2}{2m^*}  
\label{eq:effmass}
\end{align}
where the subscript ($N$ or $N+1$) refers to the corresponding
finite-$N$ quantity: e.g., $E_N$ is the total energy of $N$ particles and $T_N$ is the kinetic energy of $N$ particles. Similarly, $E^{(k)}_{N+1}$ means that the $(N+1)$-th particle is placed
at $k$: we choose several distinct values
starting near the first allowed one 
(``the Fermi surface'') and going up in magnitude as per the discrete values
in Eq.~(\ref{eq:Born}): the allowed momentum values are proportional to $\sqrt{n_x^2 + n_y^2 + n_z^2}$.
As a result, for finite $N$ the first few possible values are not very closely spaced
(on the other hand, when the integers are large, the spacing from one point to the next can be very small).
The $\xi$ is a parameter that reflects how the energy-per-particle (in the thermodynamic limit) changes in comparison 
to the non-interacting gas~\cite{Gandolfi:2015}. (Note that the number in the numerators in Eq.~(\ref{eq:effmass}) is 2, not 3).
On the right-hand side, we took the opportunity to equate to 
a single-particle energy (using $m$ for the non-interacting case and $m^*$ for the interacting case):
the left-hand side is the result of a microscopic calculation, while the right-hand side is an approximation.

As a first step, we carried out AFDMC simulations for $66$ and $67$ particles at a fixed 
density of $0.05$ fm$^{-3}$. In Fig.~\ref{fig:vsk}
we show results for $\Delta E^{(k)}_{N} $ as a function of the excitation momentum squared. 
Note that the setup of the problem in periodic boundary conditions leads to specific discrete
values for the values of $k$: as a result, there's a point ``missing'' in between our 4 points, since
there is no way to produce $\sqrt{7}$ from Eq.~(\ref{eq:Born}) (assuming a cubic box). 

Overall, the trend is quadratic (linear as a function of $k^2$), as one would expect: the whole point of the effective-mass approximation
is that one can encapsulate the complicated many-body interaction and correlation effects into a
simpler qualitative picture: the right-hand side of Eq.~(\ref{eq:effmass}) contains $m^*$ in precisely that role.
(As the excitation momentum $k$ becomes larger, it's reasonable to expect that the relevant physics
cannot be captured with a single parameter).
By fitting the points on Fig.~\ref{fig:vsk} to a straight line, we can extract the effective mass.
In this case, we find $m^*/m = 0.920 \pm 0.040$. We can already observe that 
this effective-mass ratio is less than 1: we will discuss this fact in more detail below, but for now
we merely note that other many-body approaches give $m^*/m$ values that range from much above to below 1.

Before repeating such an extraction of the effective mass,
we want to make sure our predictions for the quasiparticle dispersion can be trusted. As mentioned above, diffusion Monte Carlo methods
typically scale as $N^3$ with the particle number, so a major obstacle is the inability to simulate very large
systems. 
As a reminder, QMC methods like AFDMC can produce answers with very small statistical errors: 
for the case of neutron matter~\cite{Gandolfi:2015}, the systematic error is also under control, 
as far as bulk properties like the energy per particle are concerned. Thus, the only major uncertainty involved
is the extrapolation to the thermodynamic limit (TL), i.e., we may suffer from finite-size effects:
the quasiparticle-energy dispersion is a one-body property and may in principle behave differently from
the total energy (so $N=66$ might not be good enough). The effective-mass extraction
may or may not be impacted by such considerations, but the full microscopic prediction involved
here is for an observable, i.e., the quasiparticle energy dependence on $k$.

We recall
that for the energy-per-particle (here denoted by a bar over the relevant quantity), the standard prescription 
of how to approximate the thermodynamic limit is:
\begin{equation}
\bar{E}_{TL} = \bar{E}_{N} - \bar{T}_{N} + \bar{T}_{TL}~.
\label{eq:TL}
\end{equation}
One subtracts out the kinetic energy for the finite system and adds back in the kinetic energy of the infinite system.
Typically in neutron-star physics~\cite{Gezerlis:2008,Gezerlis:2013,Buraczynski:2016} 
the finite-size effects are dominated by the kinetic energy, so this prescription 
works very well (given that the neutron effective 
range is $r_e \approx 2.7$ fm). Note that even the quantity that appears on the left-hand side, $\bar{E}_{TL}$, may depend on $N$: 
while this is our best estimate for the thermodynamic-limit energy-per-particle, it may have been produced
using an $\bar{E}_{N}$ for too small $N$, in which case it will be a poor estimate. Something analogous
also holds in what follows.

We are now interested in applying this prescription to Eq.~(\ref{eq:effmass}),
in order to arrive at a thermodynamic-limit extrapolated quasiparticle energy dispersion relation. 
Converting from energy-per-particle to energy and taking 
the $N \rightarrow \infty$ limit in the $\bar{T}_{TL}$ term, it is not too onerous to show that 
the desired quasiparticle energy can be expressed as:
\begin{equation}
\Delta E^{(k_{TL})}_{TL} = \Delta {E}^{(k)}_{N} - \Delta {T}^{(k)}_{N} + \frac{\hbar^2 k_{TL}^2}{2m}
\label{eq:TL2}
\end{equation}
where, crucially, the last term contains the bare mass, $m$. 
It is pleasing to see how compact this new result is: in words, it says that you can convert the 
finite-$N$ quasiparticle energy to a thermodynamic-limit quasiparticle energy simply by adding
in the bare energy of the extra particle, so long as you place it at the appropriate thermodynamic-limit momentum, 
$k_{TL}$. This momentum hasn't appeared before, but turns out to be simply our old finite-size prescription,
Eq.~(\ref{eq:TL}), in a new guise:
\begin{equation}
k_{TL}^2 = k^2 - k_{F,N}^{2} + k_F^2
\label{eq:TL3}
\end{equation}
As mentioned earlier, $k_F$ is the thermodynamic limit of $k_{F,N}$.
Putting everything together, the following relation:
\begin{equation}
\Delta E^{(k_{TL})}_{TL} = \frac{\hbar^2 k_{TL}^2}{2m^*}
\label{eq:TL4}
\end{equation}
becomes the thermodynamic-limit version of Eq.~(\ref{eq:effmass}). 

\begin{figure}[t]
\begin{center}
\includegraphics[width=1.0\columnwidth,clip=]{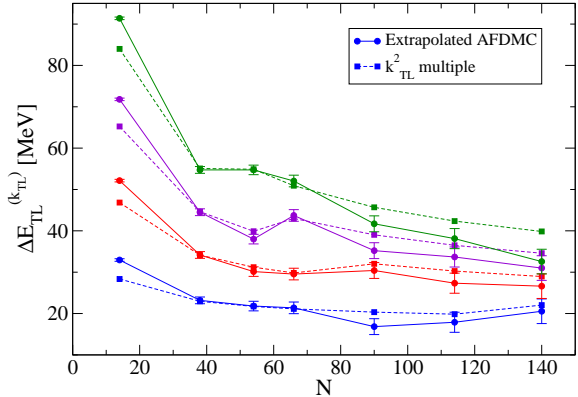}
\caption{(color online) Neutron-matter quasiparticle energy ($\Delta E^{(k_{TL})}_{TL}$)
as a function of the particle number $N$ (using AFDMC and AV8'+UIX
at a density of $0.05$ fm$^{-3}$). The solid lines 
show the first, second, third, and fourth excited-state energies (bottom to top)
following from AFDMC calculations, extrapolated to the TL.
The dotted lines show the corresponding $k_{TL}^2$ 
multiplied with a coefficient, in an attempt to match $\Delta E^{(k_{TL})}_{TL}$ as closely
as possible, see Eq.~(\ref{eq:TL4}).
\label{fig:vsN}}
\end{center}
\end{figure}

We have carried out AFDMC calculations at a density of $0.05$ fm$^{-3}$ 
for several (both closed-shell and open-shell) choices of $N$
(including its neighbor each time, at $N+1$). These have then been extrapolated to (the best-possible
approximation to) the thermodynamic limit, as per Eq.~(\ref{eq:TL2}). They are shown as the solid
lines in Fig.~\ref{fig:vsN}: the energy goes up as we increase the $k$ value at which the excited particle
is placed. This Figure also includes a separate set of curves, with a different provenance:
to produce these, we took $k_{TL}^2$ from Eq.~(\ref{eq:TL3}) and at a given excitation level (i.e., 
for the first, second, and so on solid curves) tried to find a multiplicative coefficient
that minimized the distance from the (extrapolated) AFDMC results $\Delta E^{(k_{TL})}_{TL}$. 
Interestingly, for very small particle numbers the multiples of $k_{TL}^2$
consistently underpredict the exact answers, whereas for very large particle numbers the opposite
happens. The $N$-dependence of $k_{TL}^2$ for each excited state is encoded in each dotted curve:
thus, given how difficult large-$N$ QMC calculations are, 
the expectation that ``larger $N$ means better approach to the thermodynamic limit''
is not quite met. In contradistinction to this, we find that for $N=54$ and $N=66$ the microscopic
results and the scaling results match quite well, increasing our confidence in those predictions.
Since $N=66$ outperforms $N=54$ for the closed-shell case, we consider it to be the optimal
choice.

\begin{figure}[t]
\begin{center}
\includegraphics[width=1.0\columnwidth,clip=]{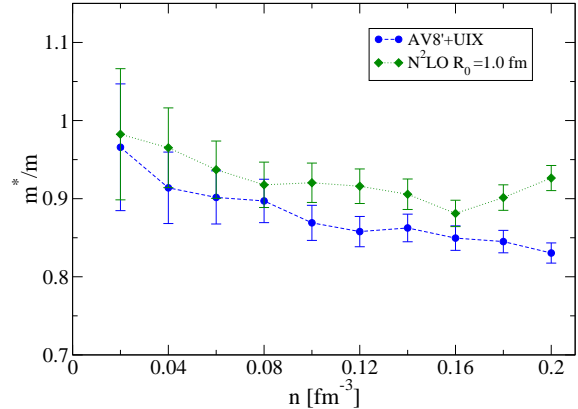}
\caption{(color online) Extracted AFDMC effective-mass ratio $m^*/m$ for neutron matter 
as a function of number density $n$. Shown are results using both phenomenological and chiral nuclear
forces. At low density the effective mass 
approaches the bare mass; as the density is increased we find a steady drop
in the value of the effective-mass ratio. 
\label{fig:vsden}}
\end{center}
\end{figure}

Having pinned down the systematic errors, we are now in a position to repeat the process of extracting
the effective mass at several densities. As mentioned above, we study densities several times smaller
than the nuclear saturation density, all the way up to $0.20$ fm$^{-3}$. 
Carrying out a straight-line fit for Eq.~(\ref{eq:TL4}), or Eq.~(\ref{eq:effmass}), and then using standard error propagation,
we can find both the effective-mass ratio $m^*/m$ and the corresponding error. The results are
shown in Fig.~\ref{fig:vsden}: our answer for $m^*/m$ is always smaller than 1 and exhibits
a decreasing trend as the density is increased, reaching a minimum value of 0.8 at our highest density.
Comparing the results stemming from the phenomenological vs the chiral nuclear potentials, 
we see that (while we have agreement at low density, as expected) when the density is increased
the qualitative behavior is similar (exhibiting a drop), but the detailed values differ.
To the degree that our QMC method is generally accurate
and the finite-size effects are under control, this is a model-independent extraction 
of the effective mass for two large classes of nuclear forces. 
In other words, our result is the product of what are known as the $k$-mass
and the $E$-mass~\cite{Li:2018}, i.e., it is the full effective-mass ratio.

The specific value of the effective mass
in neutron matter impacts static-response properties, the single-particle level spacing in neutron 
drops, as well as collective oscillations in finite nuclei.
With this in mind, let us now compare and contrast our main results with those of other theoretical 
methods~\cite{Li:2018,Boulet:2018}. Some Skyrme energy-density functionals (like SkP) employ an $m^*/m$ that is larger than 
1. Others, like SIII and SLy4 exhibit a decrease: our results seem to be intermediate between
these two functionals' $m^*/m$ ratios. Other many-body approaches typically have $m^*/m$ that first
rises above 1 and then drops back down as the density is decreased ~\cite{Wambach:1993,Schwenk:2003,Drischler:2014,Isaule:2016}. This trend is different
from what we find using our systematic non-perturbative Quantum Monte Carlo method; 
the closest similarity is with the variational calculations of Ref.~\cite{Friedman:1981}.

It's worth pointing out that extractions of the effective-mass
based on QMC calculations have been carried out for the unitary Fermi gas~\cite{Carlson:2005} 
(experimentally probed in cold-gas experiments). These give $m^*/m = 0.92$, a ratio which is comparable
to our findings at low density, but quite different from what we see near saturation density.
This is confirmation of the more complicated nature of nuclear many-body correlations, as well
as the nature of nucleon-nucleon and three-nucleon interactions. On the other
hand, the effective-mass ratio for the neutron polaron~\cite{Forbes:2014} is larger than 1,
similarly to what is found for the polaron problem at unitarity~\cite{Lobo:2006}: this reflects
the different physics involved when the effective mass studied is near the $k=0$ vs near the Fermi surface.

In conclusion, we have used particles in periodic boundary conditions to simulate the physics of 
strongly interacting matter. By introducing an extra particle at several possible excited states,
we have carried out a non-perturbative study of the quasiparticle energy dispersion of neutron matter. 
We followed a novel prescription to investigate the finite-size effects that are inherent in our
Quantum Monte Carlo formalism and made predictions for the single-particle energies. These were then used
to extract the effective-mass parameter across a large spectrum of densities, using two classes of
nuclear interactions. Our findings may in the future be incorporated in Skyrme or other nuclear
energy-density functionals, thereby improving the description of neutron-rich nuclei and neutron-star
physics.

\begin{acknowledgments}
The authors are grateful to D. Lacroix for many helpful conversations and for a careful reading of the manuscript.
They also wish to acknowledge related discussions with A. Boulet, S. Gandolfi, M. Grasso, and I. Tews. 
This work was supported
by the Natural Sciences and Engineering Research Council (NSERC) of Canada, the 
Canada Foundation for Innovation (CFI), and the Early Researcher Award (ERA) program of the Ontario Ministry of Research, 
Innovation and Science. 
Computational resources were provided by SHARCNET and 
NERSC.
\end{acknowledgments}

\end{document}